\documentclass{nature}
\usepackage[utf8]{inputenc}
\usepackage{amsmath} 
\usepackage{amssymb} 
\usepackage{amsthm} 
\usepackage{array} 
\usepackage{bm} 
\usepackage{empheq}
\usepackage{float} 
\usepackage{mathtools} 
\usepackage{mhchem} 
\usepackage{multirow}
\usepackage{physics}
\usepackage{siunitx} 
\usepackage{tensor}
\usepackage{textcomp} 
\usepackage[hyphens]{url}
\usepackage{xcolor}
\usepackage{xr-hyper}

\usepackage[final]{pdfpages} 

\usepackage{hyperref} 
\hypersetup{
    colorlinks=true,
    linkcolor=black,
    filecolor=cyan,      
    urlcolor=magenta,
    citecolor=black,
    pdftitle={Sharelatex Example},
    bookmarks=true,
    pdfpagemode=FullScreen
}

\usepackage{graphicx} 
\usepackage{subcaption} 

\usepackage{listings}
\definecolor{codegreen}{rgb}{0,0.6,0}
\definecolor{codegray}{rgb}{0.5,0.5,0.5}
\definecolor{codepurple}{rgb}{0.58,0,0.82}
\definecolor{backcolour}{rgb}{0.95,0.95,0.92}

\lstdefinestyle{mystyle}{
    backgroundcolor=\color{backcolour},   
    commentstyle=\color{codegreen},
    keywordstyle=\color{magenta},
    numberstyle=\tiny\color{codegray},
    stringstyle=\color{codepurple},
    basicstyle=\ttfamily\footnotesize,
    breakatwhitespace=false,         
    breaklines=true,                 
    captionpos=b,                    
    keepspaces=true,                 
    numbers=left,                    
    numbersep=5pt,                  
    showspaces=false,                
    showstringspaces=false,
    showtabs=false,                  
    tabsize=2
}
\usepackage{graphicx}
\usepackage{amsmath}
\usepackage{amssymb}
\usepackage{textcomp}
\usepackage{gensymb}
\usepackage{xcolor}
\usepackage{lineno}
\usepackage{soul}
\usepackage{colortbl}
\usepackage{ulem}
\usepackage[labelfont=bf]{caption}
\usepackage{multibib}
\newcites{supp}{Supplementary References}
\makeatletter
\let\saved@includegraphics\includegraphics
\AtBeginDocument{\let\includegraphics\saved@includegraphics}
\renewenvironment*{figure}{\@float{figure}}{\end@float}
\makeatother
\newcommand{\reftextit}[1]{}
\title{\large Electrostatic moir\'e potential from twisted-hBN layers}

\author{Dong Seob Kim$^{1,2}$, Roy C. Dominguez$^{3}$, Rigo Mayorga-Luna$^{3}$, Dingyi Ye$^{4}$, Jacob Embley$^{1,2}$, Tixuan Tan$^{5}$, Yue Ni$^{1,2}$, Zhida Liu$^{1,2}$, Mitchell Ford$^{3}$, Frank Y. Gao$^{1,2}$, Saba Arash$^{1,2}$, Kenji Watanabe$^{6}$, Takashi Taniguchi$^{7}$, Suenne Kim$^{8}$, Chih-Kang Shih$^{1,2}$,  Keji Lai$^{1,2}$, Wang Yao$^{5}$, Li Yang$^{4}$, Xiaoqin Li$^{1,2*}$, and Yoichi Miyahara$^{3,9*}$}

\begin{document}
\maketitle

{\renewcommand{\baselinestretch}{1.5}
\begin{affiliations}
    \item Department of Physics and Center for Complex Quantum Systems, The University of Texas at Austin, Austin, Texas, 78712, USA.
    \item Center for Dynamics and Control of Materials and Texas Materials Institute, The University of Texas at Austin, Austin, Texas, 78712, USA.
    \item Department of Physics, Texas State University, San Marcos, Texas, 78666, USA.
    \item Department of Physics, Washington University in St Louis, St. Louis, MO, 63136, USA.
    \item Department of Physics, and HKU-UCAS Joint Institute of Theoretical and Computational Physics, The University of Hong Kong, Hong Kong, China.
    \item Research Center for Electronic and Optical Materials, National Institute for Materials Science, 1-1 Namiki, Tsukuba 305-0044, Japan.
    \item Research Center for Materials Nanoarchitectonics, National Institute for Materials Science,  1-1 Namiki, Tsukuba 305-0044, Japan.
    \item Department of Photonics and Nanoelectronics, Hanyang University, Ansan 15588, South Korea.
    \item Materials Science, Engineering and Commercialization Program (MSEC), Texas State University, San Marcos, Texas, 78666, USA.

    \thanks{
  Corresponding authors:\textcolor{blue}{{yoichi.miyahara@txstate.edu}, {elaineli@physics.utexas.edu}}}
\end{affiliations}
}

{\renewcommand{\baselinestretch}{1.7}

\date{\today}
\pagebreak

\begin{abstract}
Moir\'e superlattices formed by vertically stacking van der Waals layers host a rich variety of correlated electronic phases~\cite{andrei_graphene_2020, yankowitz_van_2019, balents_superconductivity_2020, mak_semiconductor_2022} and function as novel photonic materials~\cite{wilson_excitons_2021, kennes_moire_2021, huang_excitons_2022}. The moir\'e potential of the superlattice, however, is fixed by the interlayer coupling of the stacked functional layers (e.g. graphene) and dependent on carrier types (e.g. electrons or holes) and valleys (e.g. $\Gamma$ vs. $K$). In contrast, twisted hexagonal boron nitride (hBN) layers are predicted to impose a periodic electrostatic potential that may be used to engineer the properties of an adjacent functional thin layer~\cite{zhao_universal_2021}. Here, we show that this potential is described by a simple theory of electric polarization originating from the interfacial charge redistribution, validated by its dependence on supercell sizes and distance from the twisted interfaces. We demonstrate that the potential depth and profile can be further controlled by assembling a double moir\'e structure. When the twist angles are similar at the two interfaces, the potential is deepened by adding the potential from the two twisted interfaces, reaching $\sim$ 400 meV. When the twist angles are dissimilar at the two interfaces, multi-level polarization states are observed. As an example of controlling a functional layer, we demonstrate how the electrostatic potential from a twisted hBN substrate impedes exciton diffusion in a semiconductor monolayer. These findings suggest exciting opportunities for engineering properties of an adjacent functional layer using the surface potential of a twisted hBN substrate. 
\end{abstract}
}

\maketitle
\renewcommand{\baselinestretch}{2}
\setlength{\parskip}{7pt}
Moir\'e superlattices have proven to be an exceptionally rich material platform in which different phases of matter can be realized by engineering electronic bands via either changing the twist angle, doping, or electric field~\cite{andrei_graphene_2020, mak_semiconductor_2022, wilson_excitons_2021, huang_excitons_2022}. The existing moir\'e  systems, however, have limitations in various aspects. Firstly, moir\'e  patterns useful for electronic band engineering typically form between van der Waals (vdW) layers with similar lattice constants. Secondly, the potential depth is fixed by interlayer coupling and not easily adjustable. Thirdly, lattice and electronic properties are necessarily coupled, resulting in  undesirable constraints such as lattice reconstructions at small twist angles.  Thus, a new approach that separates the generation of the moir\'e potential from the functional layer would significantly expand the flexibility in moir\'e engineering.

Hexagonal boron nitride (hBN) as a wide-gap insulator has played an essential role in van der Waals (vdW) materials and heterostructures. In the vast majority of the studies, hBN layers act as passive layers such as atomically smooth substrates~\cite{xue_scanning_2011}, capping layers~\cite{lee_highly_2015}, or ultrathin tunnel barriers~\cite{britnell_electron_2012}, drastically improving charge carrier mobility~\cite{dean_boron_2010} or reducing inhomogeneous broadening of optical resonances. One exception is that moir\'e patterns may form between hBN and graphene layers due to their similar lattice constants, modifying the properties of the multilayer~\cite{wang_composite_2019}. hBN itself offers superior thermal conductivity and interesting photonic and phononic properties~\cite{caldwell_photonics_2019}. For example, hBN exhibits natural hyperbolic dispersion in the mid-infrared range~\cite{li_hyperbolic_2015}, hosts defect-bound single photon emitters at room temperature~\cite{tran_quantum_2016}, and functions as UV photon detectors~\cite{maity_hexagonal_2018}. Very recently, twisted hBN bilayers have been found to exhibit ferroelectric states in both transport and scanning probe measurements~\cite{yasuda_stacking-engineered_2021, vizner_stern_interfacial_2021, woods_charge-polarized_2021, moore_nanoscale_2021, chiodini_moire_2022}. 

Here, we demonstrate that the electrostatic potential on the surface of a twisted hBN (t-hBN) bilayer or a multilayer can be used to impose a universal moir\'e potential on an adjacent functional layer. This moir\'e potential is tunable in several ways. First, the potential depth changes with the supercell size and the top hBN layer thickness. The magnitude of the potential agrees with predictions from a simple theory of electric polarization originating from the interfacial charge redistribution. Furthermore, this surface potential can be engineered in a double moir\'e configuration formed by three twisted hBN layers. When the supercell sizes of the two moir\'e patterns are similar, the potential at each interface adds constructively, leading to a deeper potential modulation of $\sim$ 400 meV. When the supercell sizes of the two moir\'e patterns are different, multiple polarization states form, enabling different applications, i.e. ferroelectric domains for multi-state memory. As an example of controlling optical properties of a functional layer, we demonstrate how a twisted hBN substrate can impede exciton diffusion in an adjacent MoSe$_2$ monolayer. Our work may stimulate future studies that combine the t-hBN substrates with rather different materials (e.g. layered materials with different lattice constants and symmetry or polar molecules and polymers), thus, expanding the footprint of moir\'e engineering in materials science.

\begin{figure}[H]
    \centering
    \includegraphics[width=16.5cm]{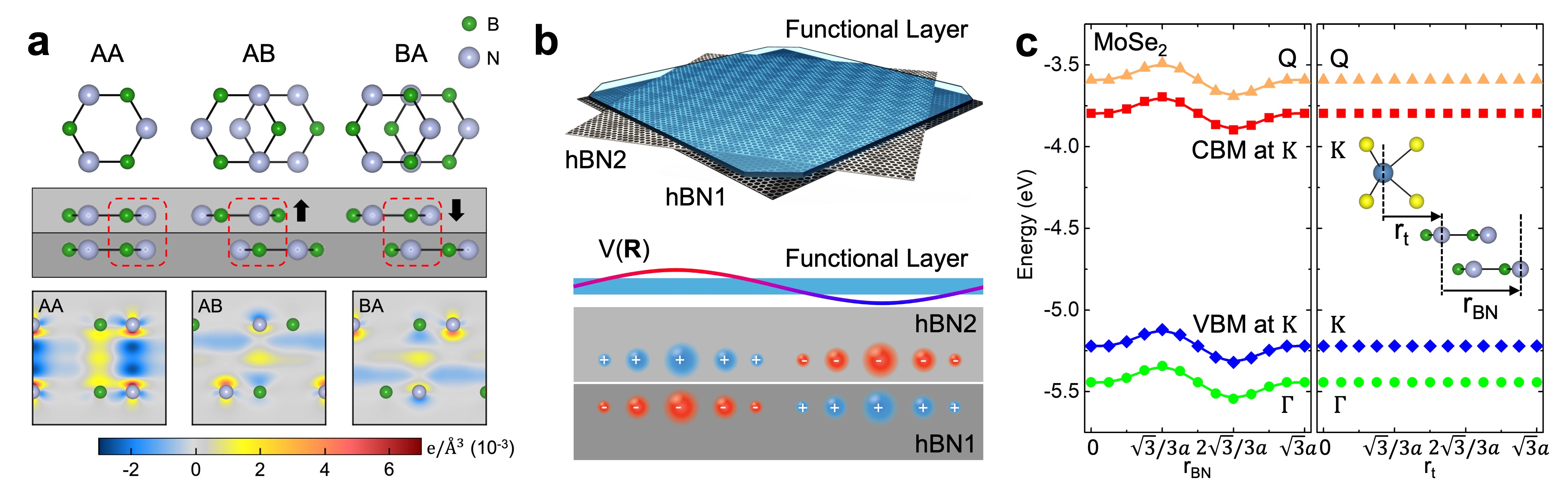}
    \caption{\textbf{Electrostatic moir\'e potential from a t-hBN substrate modifies properties of an adjacent functional layer}
    \textbf{a}, Top and side views of the hBN parallel interface at high-symmetry points. Calculated charge redistribution corresponding to the three dashed red boxes. Black arrow represents a net polarization. 
    \textbf{b}, Illustration of an electrostatic moir\'e potential V(\textbf{R}) at the top surface of a t-hBN substrate that modifies the properties of an adjacent functional layer.
    \textbf{c}, modulation of band-edge energies in MoSe$_2$ by a t-hBN substrate. Atomic registry between the two hBN layers (r$_{BN}$) modulates the conduction band minima and valence band maxima (left). In contrast, atomic registry between MoSe$_2$ and t-hBN does not induce change in the band (right).}
    \label{fig:fig1}
\end{figure}

We first describe the formation of a net polarization and the electrostatic potential of a t-hBN bilayer conceptually as shown in Fig.~\ref{fig:fig1}a. Commonly used hBN crystals are naturally stacked in an AA' sequence, i.e., B atoms are vertically aligned with N atoms in adjacent layers. In another energetically favorable AB or BA (Bernal) stacking configuration, a perpendicular electric polarization emerges because both inversion and out-of-plane symmetry are lifted at these high symmetry points, unlike in the case of AA' stacking. An electric polarization (indicated by black arrows) originates from charge redistribution $\Delta\rho$ at the buried interface (Fig.~\ref{fig:fig1}a). The Bernal stacking can be realized as a t-hBN bilayer with a nearly parallel interface. In this case, the electrostatic potential is described by
\begin{equation}
    V(\mathbf{R}, z)
    \approx \text{sgn}(z)\frac{P(\mathbf{R})}{2\epsilon_{0}}e^{-G|z|},
    \label{eqn:eqn1}
\end{equation}
where the net polarization is calculated from $P(\mathbf{R})=\int z'\Delta\rho(\mathbf{R},z')dz'$, $G=\frac{4\pi}{\sqrt{3}b}$, $\mathbf{R}$ represents the lateral position vector, $b$ characterizes the supercell size, and $z$ is the vertical distance to the buried interface~\cite{zhao_universal_2021}. The magnitude of $P$ is 2.01 pCm$^{-1}$ obtained from the first principles calculations (Fig.~\ref{fig:fig1}a), consistent with previous experiments~\cite{yasuda_stacking-engineered_2021, vizner_stern_interfacial_2021, woods_charge-polarized_2021}. In a t-hBN bilayer, the lateral variation of this potential $V(\mathbf{R})$ follows the superlattice periodicity, and its amplitude reduces with decreasing supercell sizes.

The moir\'e potential generated by a t-hBN substrate may modulate the electronic bands of an adjacent thin functional layer as illustrated by Fig.~\ref{fig:fig1}b. As an example, we calculate a semiconducting MoSe$_2$ monolayer placed on two hBN monolayers with a parallel interface. The orientations of crystalline axes of MoSe$_2$ and hBN layers are assumed to be aligned with a 4:3 ratio of their lattice constants in the calculation. The conduction and valence band edge states are modulated as the atomic registry of the t-hBN (r$_{BN}$) varies as shown in the left panel of Fig.~\ref{fig:fig1}c. This calculation confirms that all band-edge states are modulated with the same profile as a function of r$_{BN}$, which is one reason for the term ``universal potential". In contrast, translating the MoSe$_2$ monolayer on top of the t-hBN substrate (e.g. as a function of r$_t$) does not influence the bands of the functional layer, making this remote potential modulation a robust approach for band engineering.

\begin{figure}[H]
    \centering
    \includegraphics[width=13cm]{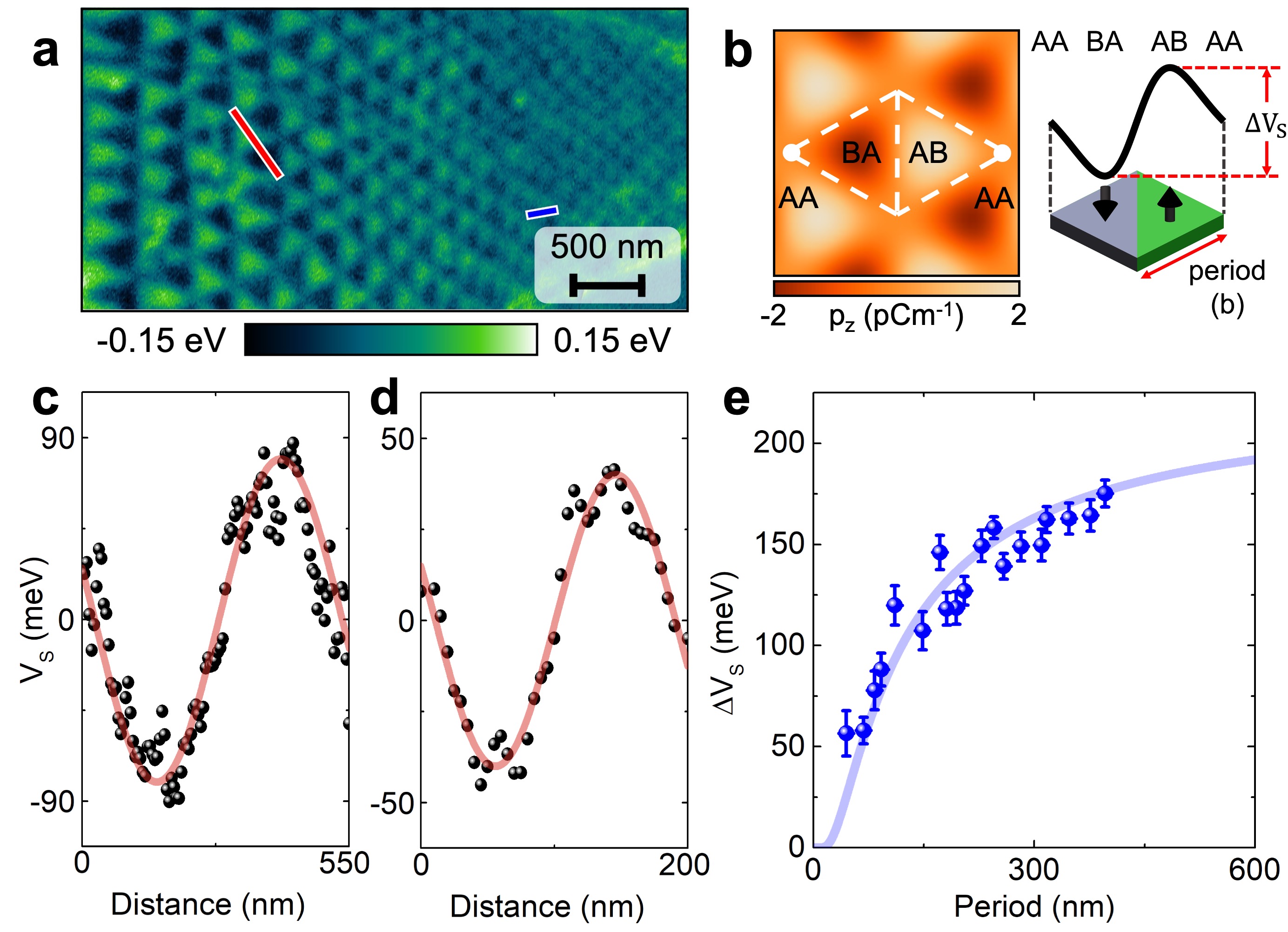}
    \caption{\textbf{Supercell size dependence of the moir\'e potential at the surface of a t-hBN bilayer.}
    \textbf{a}, A representative KPFM image showing supercell sizes changing gradually. 
    \textbf{b}, Calculated real space polarization map for a supercell (left). Illustration of the surface potential modulation depth, $\Delta V_S$, and moir\'e period, b (right).
    \textbf{c-d} Potential line profiles extracted from two different supercells (red and blue lines in panel a). The red solid lines are fitted with $A \sin{B(x-C)}$. The potential depth becomes shallower for a smaller supercell.
    \textbf{e}, Summary of potential depth ($\Delta V_{S}= V_{S, max} - V_{S, min}$) as a function of the moir\'e period b. The potential depth saturates as the supercell size increases. The solid blue line is the theoretical curve calculated from equation~\ref{eqn:eqn1} with the measured top hBN layer thickness $z = 7.8$~nm.}
    \label{fig:fig2}
\end{figure}

We now experimentally quantify the electrostatic potential at the surface of the t-hBN substrate. The t-hBN layers are assembled by folding as illustrated in Extended Data Fig.~\ref{fig:figs1} (more details in Methods). When the hBN is folded along the armchair (zigzag) direction, a parallel (anti-parallel) interface forms as shown in recent experiments~\cite{yasuda_stacking-engineered_2021, vizner_stern_interfacial_2021, woods_charge-polarized_2021}. 
Self-folding commonly leads to t-hBN bilayers with a marginal twist angle~\cite{chang_graphene_2018}. The resulting moir\'e superlattice can be imaged with several scanning probe techniques and transmission electron microscopy~\cite{jiang_electron_2018, shabani_deep_2021}. 
Here, we measure the electrostatic potential at the top surface of the t-hBN layers using frequency modulation Kelvin Probe Force Microscopy (KPFM) (details in Methods)~\cite{zerweck_accuracy_2005}. 
A representative KPFM image is shown in Fig.~\ref{fig:fig2}a. A gradual change in the supercell size is observed and likely originates from strain-induced twist-angle variations. 
The supercell sizes range from $\sim$ 500~nm to $\sim$ 50~nm. The twist angles are estimated to be between 0.03$^\circ$ and 0.3$^\circ$ using the relation $b = a/\delta$ where $b$ is the supercell size, $a$ is the hBN lattice constant, and $\delta$ is the twist angle. The color contrast in the KPFM image corresponds to the surface potential.

We perform detailed analysis of the supercell size (i.e. twist-angle) dependence of the surface potential in Fig.~\ref{fig:fig2}. We anticipate the measured potential to be dominated by the leading Fourier harmonics expansion, i.e., a sinusoidal function~\cite{vizner_stern_interfacial_2021} where the maximal and minimal values occur at AB and BA points. The first principles calculations and the illustration supporting this expected potential profile are shown in Fig.~\ref{fig:fig2}b. The line profiles along the diagonal directions (labeled as red and blue lines in Fig.~\ref{fig:figs2}a) are presented in Fig.~\ref{fig:figs2}c-d, in which the solid line is a sinusoidal fitting function, $A\sin{B(x-C)}$. The depth of the potential $\Delta V_{S}= V_{S, max} - V_{S, min}$ extracted from the fitting is plotted as a function of the supercell size in Fig.~\ref{fig:fig2}e. The error bars along the horizontal and vertical-axes indicate the uncertainty of supercell size and surface potential, respectively. The potential increases and becomes saturated with increasing supercell sizes. This observation is in excellent agreement with the prediction (solid blue line) of equation~\ref{eqn:eqn1} taking into account the measured top hBN thickness of $z = 7.8$~nm. In plotting these theoretical curves, the maximum absolute value of the potential is set to be 225~meV unless stated otherwise. In all measurements, the tip of the KPFM is kept at $\sim$ 2~nm above the sample. The tip-to-sample distance dependence is further discussed in the Extended Data Fig.~\ref{fig:figs2}. In the analysis, we chose regions of the sample with equilateral triangle patterns, which are less susceptible to unintentional extrinsic strain, to extract the potential value. The supercell shape influences the extracted potential modulation as further discussed in Extended Data Fig.~\ref{fig:figs3}.

\begin{figure}[H]
    \centering
    \includegraphics[width=11.5cm]{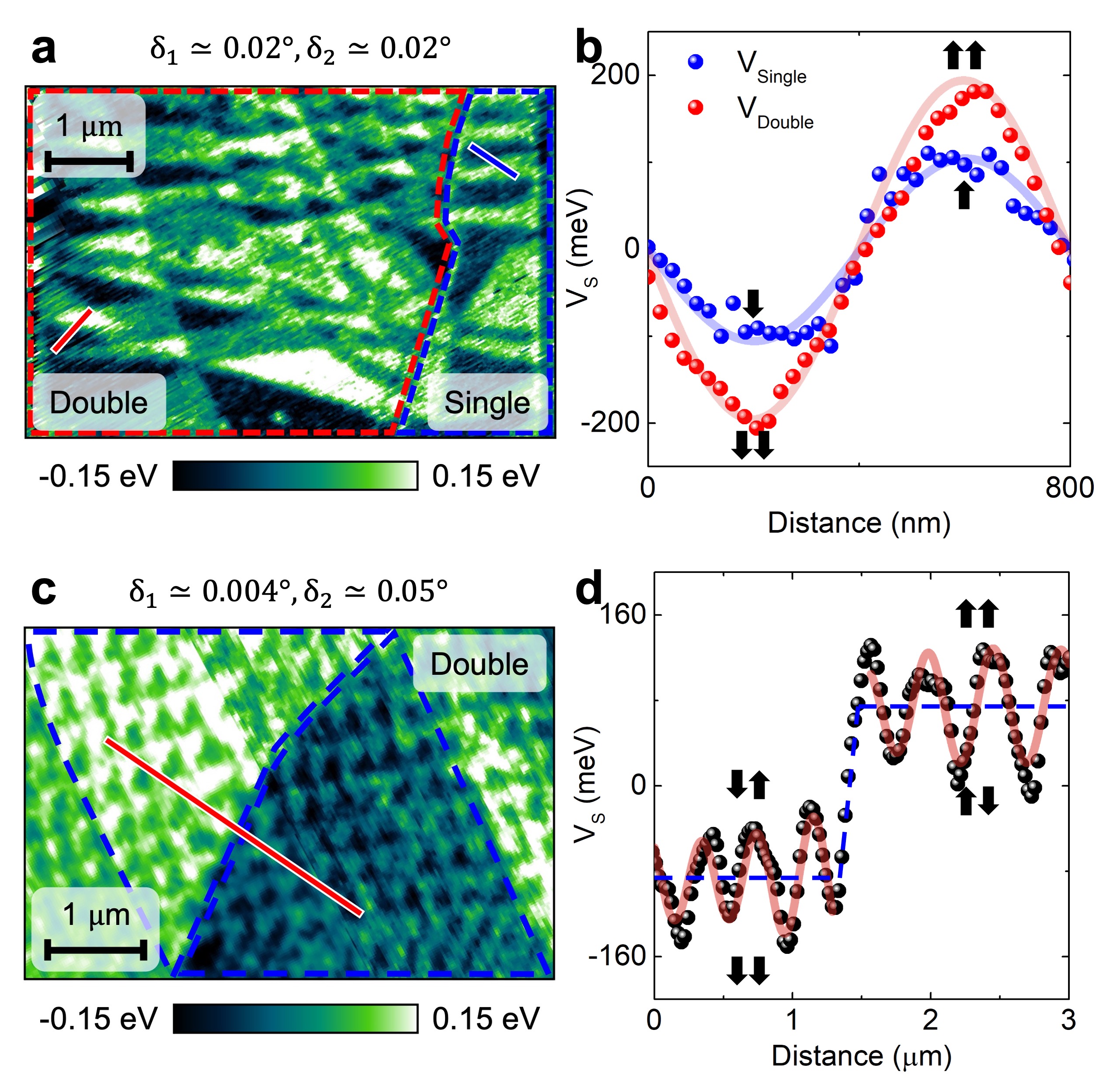}
    \caption{\textbf{Enhanced potential depth and multi-level polarization states in double moir\'e superlattices and multilayers.}
    \textbf{a} A KPFM image of single (dashed blue) and double (dashed red) moir\'e regions. The twist angles at both interfaces are 0.02$\degree$ in the double moir\'e.
    \textbf{b}, Comparing potential modulations from supercells with similar sizes in single and double moir\'e regions. The line profiles are extracted along the blue and red lines in \textbf{a}, respectively. The solid lines are sinusoidal fittings. Electrostatic potentials at two interfaces of the double moir\'e add and deepen the potential depth. Black arrows indicate polarization directions.
    \textbf{c}, KPFM image of a double moir\'e superlattice with dissimilar twist angles at both interfaces. Small supercells of $\sim$ 300 nm superimpose on large supercells of $3.5~\mu$m encircled by dashed blue lines. 
    \textbf{d}, A line profile along the red line marked in (\textbf{c}). Multi-level states correspond to regions from the same or opposite polarizations (black arrows) at the two interfaces. Solid red line is sinusoidal fittings and dashed blue line is a guide to the eye.}
    \label{fig:fig3}
\end{figure}

Having established a good understanding of t-hBN bilayers, we further engineer the moir\'e potential by assembling double moir\'e structures consisting of three t-hBN layers. In the example shown in Fig.~\ref{fig:fig3}a, three hBN layers each with a thickness of $\sim 20$~nm and twist angles $\simeq0.02\degree$ are assembled. The surface potentials from the single and double moir\'e regions are compared by taking two line profiles plotted in Fig.~\ref{fig:figs3}b from regions indicated by the blue and red lines indicated in Fig.~\ref{fig:fig3}a.) The potential modulation in the double moir\'e structure (red data points and curve) in Fig.~\ref{fig:fig3}b is nearly twice of that in the single moir\'e superlattice (blue data points and curve) when two supercells of $\sim$ 800~nm are analyzed. In the double moir\'e region, the potential modulation is as deep as $V_{double}$= 390~meV (Fig.~\ref{fig:fig3}b) because the electrostatic potential from each interface adds constructively. In this example, we analyzed large supercells approaching $\sim$ 1 $\mu$m. Deviations from the sinusoidal function due to lattice reconstructions are observed near the peaks, e.g. data points taken in the single moir\'e region (blue points). Complete lattice relaxations of moir\'e superlattices would lead to domains with AB and BA atomic registry separated by sharp domain walls. 

Another example of the double moir\'e structure is presented in Fig.~\ref{fig:fig3}c, which consists of two interfaces with very different periods of $\sim$ 3.5~$\mu$m and $\sim$ 300~nm, corresponding to twisted angles of $0.004\degree$ and $0.05\degree$, respectively. Each layer has a thickness of $\sim 20$~nm. Stacked triangular supercells are clearly observed where the larger supercell is encircled by a dashed blue line. Taking a line cut (red line) of the potential modulation across the boundary of two large supercells, we observe a step-like increase of $\sim 150$~mV superimposed on a sinusoidal modulation with an amplitude of $\sim 120$~mV (Fig.~\ref{fig:fig3}d). The solid line is a sinusoidal fitting function while the dashed line is a guide to the eye. This pattern corresponds to the formation of two polarization states in this double moir\'e structure. The accumulative polarization from multiple interfaces has been recently reported in transition metal dichalcogenide multilayers~\cite{deb2022cumulative}, consistent with our findings.

\begin{figure}[H]
    \centering
    \includegraphics[width=13cm]{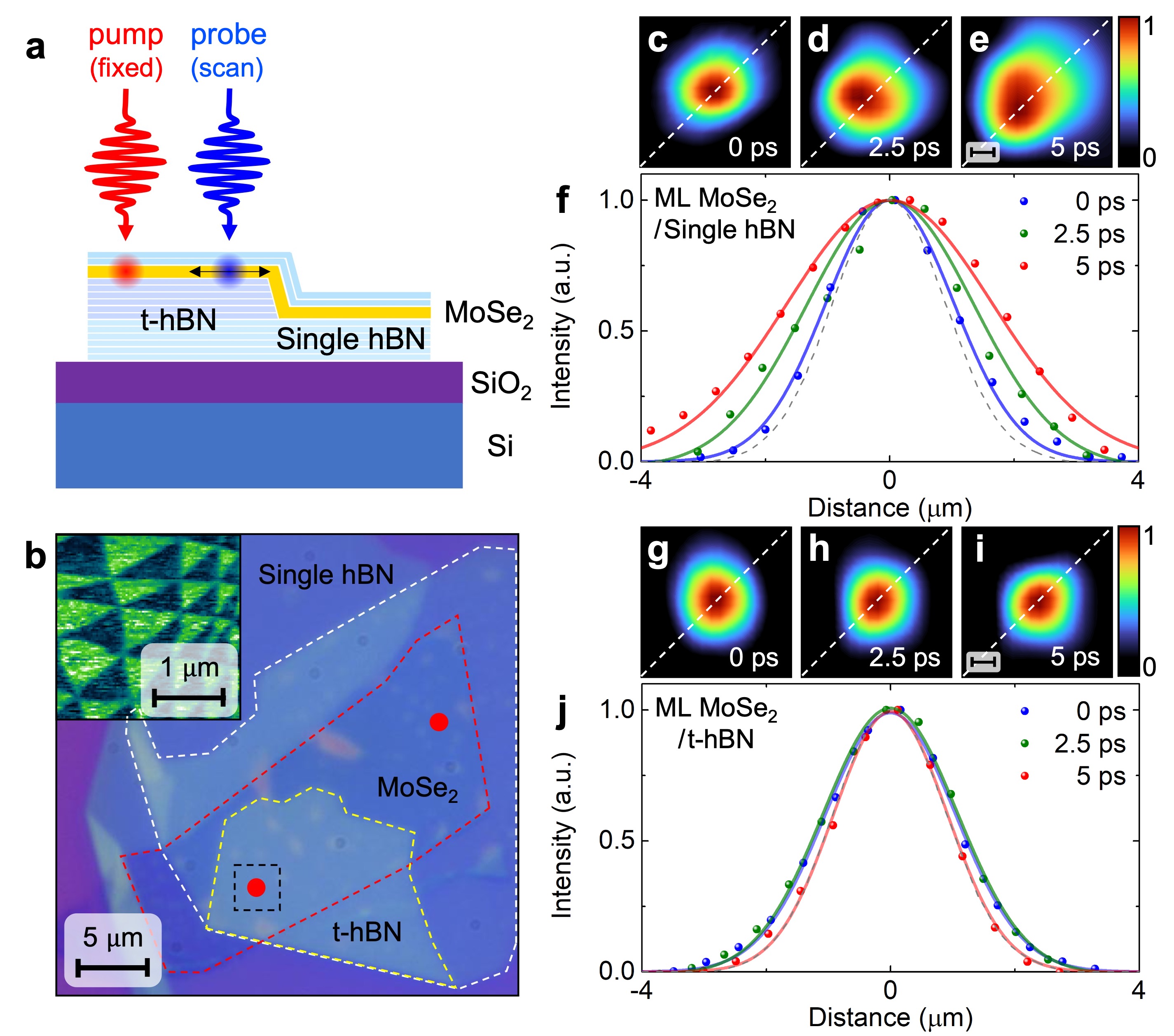}
    \caption{\textbf{Moir\'e potential from a t-hBN substrate impedes exciton diffusion in a monolayer MoSe$_2$.}
    \textbf{a}, Schematic of the MoSe$_2$/t-hBN multilayer and the spatially-resolved exciton diffusion measurement. Probe beam is raster scanned across an area centered around a fixed pump beam.
    \textbf{b}, Optical microscope image of the monolayer MoSe$_2$ placed on either a single hBN or a t-hBN bilayer. Inset: KPFM image of a region on the t-hBN bilayer where optical measurements are taken.
    \textbf{c-e}, Normalized pump-probe images on MoSe$_2$ monolayer (ML) on top of a single hBN at delay times of 0 ps (\textbf{c}), 2.5 ps (\textbf{d}), and 5 ps (\textbf{e}). The scale bar is 1 $\mu$m.
    \textbf{f}, Line profiles taken along the white dashed lines in (\textbf{c-e}).
    \textbf{g-j}, Corresponding images and line profiles from the ML MoSe$_2$ placed on top of the t-hBN substrate.}
    \label{fig:fig4}
\end{figure}

Finally, we demonstrate how the electrostatic potential generated by a t-hBN bilayer can impede exciton diffusion in a MoSe$_2$ monolayer. As a prototypical vdW semiconductor, optical properties of MoSe$_2$ monolayers and heterostructures have been investigated extensively. For example, exciton lifetimes and diffusion are both critical properties in determining the performance of optoelectronic devices, and they are modified drastically by moir\'e potential in heterostructures, e.g. MoSe$_2$/WSe$_2$~\cite{huang_excitons_2022}. Here, we show that diffusion of intralayer excitons in a MoSe$_2$ monolayer is impeded by the moir\'e potential imposed by a t-hBN substrate. The layered structure and its optical microscope image are shown in Fig.~\ref{fig:fig4}a-b. We perform spatially-resolved pump-probe experiments to compare exciton diffusion in two regions (marked by red dots in Fig.~\ref{fig:fig4}b) of the sample with either a single hBN layer or a t-hBN bilayer with supercells of $\sim$ 700 nm. The measurements are taken at 10 K, and the wavelength of both pump and probe is tuned to the A exciton (more details in Method).

The spatial images of exciton diffusion taken from the MoSe$_2$/single hBN region are displayed in Fig.~\ref{fig:fig4}c-e while the line cuts (white dashed lines) from each image are shown in Fig.~\ref{fig:fig4}f. As the delay time between the pump and probe increases, exciton diffusion beyond the excitation laser spot (black dashed curve) is clearly observed. In contrast, no exciton diffusion is observable from the MoSe$_2$/t-hBN region (Fig.~\ref{fig:fig4}g-j). Similar data taken from different locations on the sample and at different exciton densities are shown in the Extended Data Figs.~\ref{fig:figs9} and ~\ref{fig:figs10}.  In many previous studies of TMD monolayers encapsulated by hBN, exciton diffusion was observed~\cite{wagner2021nonclassical}. These experiments demonstrate that electrostatic potential from the t-hBN substrate functions differently from encapsulation layers and effectively impedes exciton diffusion in the MoSe$_2$ monolayer. Mobile carriers (electrons or charged excitons) in two-dimensional systems (quantum wells and TMD monolayers) can be confined or completely localized by lateral electrostatic potential. The mechanisms impeding exciton diffusion in the current experiments are complex and need further studies. One possibility is that excitons are polarized and trapped by the potential. The band structures of the TMD monolayers are modified by the remote moir\'e potential, leading to the emergence of flat bands. When the size of lateral confinement (5-20 nm) is on the same order as exciton Bohr radius (1-2 nm in TMDs), carriers may be completely localized by the moir\'e potential, which essentially defines a regular array of quantum dots (zero-dimensional quantum systems). Near-field optical techniques with higher spatial resolutions would be necessary to demonstrate neutral or charged exciton localization.

We compare the experiments presented here with other recent studies of t-hBN layers. It has been recognized that spontaneous polarizations form at the interface of t-hBN layers and ferroelectric domains can be controlled via a bias voltage~\cite{vizner_stern_interfacial_2021}. However, the electrostatic  potential has never been studied systematically, preventing the evaluation of t-hBN layers as a general approach for moir\'e engineering. One previous experiment extracted the surface potential using electrostatic force spectroscopy ~\cite{woods_charge-polarized_2021}.
In their measurements, the tip oscillation phase is measured as a function of applied voltage at the centers of two neighboring domains. They measured the shift of the parabola maxima of the resulting phase vs voltage curves from which the potential depth was extracted. This method of extracting the potential depth relies on the measurements at only two locations and, thus, is more susceptible to influences of measurement noises or surface adsorbates, especially for smaller supercells. In addition, we found supercells are often distorted in shape, e.g. exhibiting elongation along one direction. The surface potential from such distorted supercells shows a deeper potential along the elongated direction, making a comparison between different supercell sizes challenging (Extended Data Fig.~\ref{fig:figs3}). Thus, no accurate theory-experimental comparison of the surface potential of the t-hBN layers have been reported previously.

In conclusion, we quantify the deep electrostatic potential on the surface of t-hBN layers and suggest that it can impose a universal moir\'e potential modulation on an adjacent functional layer. Having established a simple model describing this  electrostatic potential via its systematic changes with the twist angle and distance to the interface $z$ (Extended Data Fig.~\ref{fig:figs4}), we demonstrate the accumulative effect from two interfaces in double moir\'e structures, reaching a potential as deep as $\sim$400 meV. Such a strong moir\'e potential depth has been inaccessible in previously studied moir\'e systems. As a simple example of controlling properties of an adjacent semiconductor monolayer, we show exciton diffusion is impeded by a t-hBN substrate, complementing a prior study of dielectric modulation of exciton resonances by a graphene/hBN substrate~\cite{xu_creation_2021}. Furthermore, we calculate how a combined t-hBN substrate and a gate voltage can be used to tune the electronic band of a natural graphene bilayer from topologically trivial to non-trivial (Extended Data Fig.~\ref{fig:figs6}).  The hetero-interface between the hBN and functional layer should not change the energy band of the functional layer, evidenced by the common practice of using hBN as a substrate and capping layer. Thus, our work suggests a promising approach to broaden moir\'e engineering to periodically modulate properties of a wide range of electronic and photonic functional layers.\\

\section*{Methods}

\subsection{Sample preparation:} 
We exfoliated hBN flakes using scotch tape onto 285~nm SiO$_2$/Si substrates. After choosing a target hBN flake via microscope imaging, nitrogen gas is blown to facilitate the folding process. Samples typically fold along either zigzag or armchair directions. After stacking or folding, the samples are annealed up to 500$\degree$C for 4 hours under vacuum $\sim$ 10$^{-7}$ Torr to increase interface bonding by removing polymer residue. For the MoSe$_2$/t-hBN structure, the MoSe$_2$ monolayer was exfoliated and then dry transferred using a 15\% PPC solution, dissolved into anisole.

\subsection{KPFM measurements:}
Kelvin probe force microscopy (KPFM) measurements were performed using SmartSPM (AIST-NT) in two-pass frequency modulation KPFM (FM-KPFM) mode and amplitude modulation KPFM (AM-KPFM) mode. All the data except for the one in Extended Data Fig.~\ref{fig:figs5}a were taken by FM-KPFM mode. We used Pt coated conductive cantilever probes with a nominal resonance frequency of 70~kHz and a spring constant of 2~N/m (OPUS 240AC-PP, Mikromasch) and the gold coated cantilevers with supersharp diamond-like carbon tips with a nominal resonance frequency of 150~kHz and a spring constant of 5~N/m (BudgetSensors SHR150). In the FM-KPFM mode of SmartSPM, the resonance frequency shift of the mechanically excited oscillations (amplitude 20~nm), which are caused by the electrostatic force gradient with respect to the tip-sample distance, is detected via the phase of the cantilever oscillations. 
The amplitude of the modulation of the phase, which is caused by applying an ac voltage (3~V, 1~kHz), is proportional to the difference between the applied dc bias voltage between the contact potential difference (CPD) and is fed into a feedback controller to nullify the electrostatic force. 
In FM-KPFM measurements, the resonance frequency shift is determined by the electrostatic force gradient with respect to the tip-sample distance rather than the force itself~\cite{Miyahara2017b, axt2018}. Due to the spatial averaging effect caused by the cantilever, AM-KPFM measurements typically underestimate the potential modulation up to a fifth of FM-KPFM (Extended Data Fig.~\ref{fig:figs5}). 

\subsection{Transient differential reflection spectroscopy:}
For the pump-probe measurements, 80 fs pulses are derived from a Ti:sapphire oscillator operating at a 100 MHz repetition rate. The pump beam was modulated at 100 kHz using an acoustic optical modulator for the lock-in detection. Both pump and probe beams were spatially filtered before focused to a spot size of $\sim$ 2 $\mu$m in diameter using a x50 (NA = 0.55) objective. The reflected probe beam was collected by the same objective and detected by an avalanche Si photodiode. A galvanometer scanner was used to scan the probe beam to obtain the exciton diffusion images presented in Fig.~\ref{fig:fig4}. The excitation power is 20 $\mu$W, corresponding to an exciton density of $\sim$ 7$\times$10$^{11}$ cm$^{-2}$

\subsection{Computational methods:}
First-principles calculations based on density functional theory are performed within the general gradient approximate (GGA) using the Perdew-Burke-Ernzerhof (PBE) exchange-correlation functional as implemented in the Vienna Ab initio Simulation Package (VASP)~\cite{kresse_efficiency_1996, kresse_efficient_1996}. The cut-off energy for plane wave-basis expansion is 450 eV. The vdW interlayer interactions are included via the semi-empirical Grimme-D3 scheme. The Brillouin zone is sampled with 9$\times$9$\times$1 k points using the Monkhorst-pack scheme. The out-of-plane polarization of bilayer h-BN is calculated based on the charge distribution integration and checked with geometric quantum phase (Berry phase) approach~\cite{king-smith_theory_1993} with a relaxed lattice constant a = 2.51 {\AA}. To avoid interactions between neighboring units, a 20 {\AA} vacuum space is added between bilayer units. Dipole correction is employed in the simulation.
\\
The band structure of the bilayer graphene under moir\'e potential is calculated using the following real-space Hamiltonian of the AB-stacked bilayer graphene~\cite{ghorashi_topological_2023}. We only consider electrons from the K valley.
\begin{equation}
H_{BLG}(\vec{R})=\hbar v_F \tau^0(-i\partial_x\sigma^1-i\partial_y\sigma^2)+\frac{t_{int}}{2}(\tau^1\sigma^1-\tau^2\sigma^2)
\end{equation}
$v_F$ is the Fermi velocity. $\tau$ and $\sigma$ are Pauli matrices acting on layer and sublattice spaces. $t_{int}\approx 0.33$ eV is the interlayer coupling for bilayer graphene. The applied bias and moir\'e potential generated by twisted hBN takes the following form \cite{zhao_universal_2021}\par
\begin{equation}
H_{bias}=V_0\tau^3\sigma^0
\end{equation}
\begin{equation}
H_{MP}(\vec{R})=V_{MP}((\tau^0+\tau^3)+\alpha(\tau^0-\tau^3))\sigma^0\sum_{n=1,3,5}\sin(\vec{G}_n\cdot \vec{R})
\end{equation}
According to the moir\'e potential formula given in the main text, we take $V_{MP}=C \exp(-4\pi \frac{z}{\sqrt{3}b})$ and $\alpha=\exp(-\frac{4\pi d}{\sqrt{3}})$. $d = 0.35$~nm is the interlayer distance of bilayer graphene. For Extended Data Fig.~\ref{fig:figs6}, P = 35.49 meV and z = 0.58 nm. $\vec{G}_n=\frac{4\pi}{\sqrt{3}b}(\cos(n\frac{\pi}{3}), \sin(n\frac{\pi}{3}))$

\subsection{Modulation potential fitting:}
Lattice reconstructions may occur in moir\'e superlattices with small twist angles and fall into three different regimes~\cite{quan_phonon_2021}. In the fully relaxed regime with the smallest twist angle, the superlattice is triangular and the AB (BA) domains are separated by sharp domain walls with AA stacking. In the transition regime with increasing twist angles, domains are less defined, variations between atomic alignment occur driven by intrinsic strain. In the rigid regime with large twist angles, the two layers are largely decoupled and no atomic displacements occur within each layer. The smooth variations of the electric potential shown in KPFM images may originate from either partial lattice relaxation in the transition regime. In the largest supercells, a plateaus is observed in the potential line profiles due to lattice relaxation, deviating from a sinusoidal fitting function.

\section*{References}
\bibliography{hBN-KPFM}

\section*{Data availability}
The data that support the plots within this paper and other findings of this study are available from the corresponding authors upon reasonable request. Source data are provided with this paper.

\section*{Acknowledgements} The experiments are primarily supported by the Air Force Office of Scientific Research under award number FA2386-21-1-4067 (D. S. K.), NSF ECCS-2130552 (Z. L.), and the Department of Energy, Office of Basic Energy Sciences under grant DE-SC0019398 (Y.N.). The work by R.M-L. and R.C.D. is supported by the NSF Partnership for Research and Education in Materials (PREM) (NSF award DMR-2122041). 
Y.M. acknowledges the support by the NSF CAREER (DMR-2044920)and NSF MRI (DMR-2117438) grants. X. L. gratefully acknowledges the Welch Foundation grant F-1662 for support in sample preparation. The collaboration between UT-Austin (S. A.) and Washington University was enabled by NSF Designing Materials to Revolutionize and Engineer our Future (DMREF) program via grants DMR-2118806 and DMR-2118779. 
The collaboration between X. L., C. S, and K. L are enabled by the National Science Foundation through the Center for Dynamics and Control of Materials: an NSF MRSEC under Cooperative Agreement No. DMR-1720595, which also supported the user facility where part of the experiments were performed. The Major Research Instrumentation (MRI) program DMR-2019130 (F. G.) supported some experiments. K.W. and T.T. acknowledge support from the JSPS KAKENHI (Grant Numbers 20H00354, 21H05233 and 23H02052) and World Premier International Research Center Initiative (WPI), MEXT, Japan. W.Y. acknowledges support by RGC of HKSAR (HKU SRFS2122-7S05), and the Croucher Foundation.

\section*{Author contributions}
D.S.K., X. L., and Y.M. conceived the experiment. D.S.K. fabricated samples with assistance from Y.N., Z.L., F.Y.G., and S.A. for sample characterizations. D.S.K., R.M-L. and R.C.D. carried out AFM and KPFM experiments with contributions from M.F.. D.S.K. performed optical measurements with contributions from J.E.. K.W. and T.T. synthesized the bulk crystals. D.Y., T.T., W.Y., and L.Y. proposed the theoretical model and performed the first-principles calculation. D.S.K. analyzed the data with contributions from R.M.L. and D.Y.. D.S.K., S.K., C.K.S., K.L., W.Y., L.Y., Y.M., and X.L. wrote the first draft of the manuscript. All authors contributed to discussions.

\section*{Competing interests}
The authors declare no competing interests.

\newpage
\section*{Figures}

\renewcommand{\baselinestretch}{1}

\bibdata{hBN-KPFM.bib}


\setcounter{figure}{0}
\setcounter{table}{0}

\newpage
\section*{Extended Data}

\makeatletter
\let\saved@includegraphics\includegraphics
\renewenvironment*{figure}{\@float{figure}}{\end@float}
\renewenvironment*{table}{\@float{table}}{\end@float}

\renewcommand{\figurename}{Extended Data Fig.}
\renewcommand{\tablename}{Extended Data Table.}

\renewcommand{\baselinestretch}{1}

\begin{figure}[H]
    \centering
    \includegraphics[width=10cm]{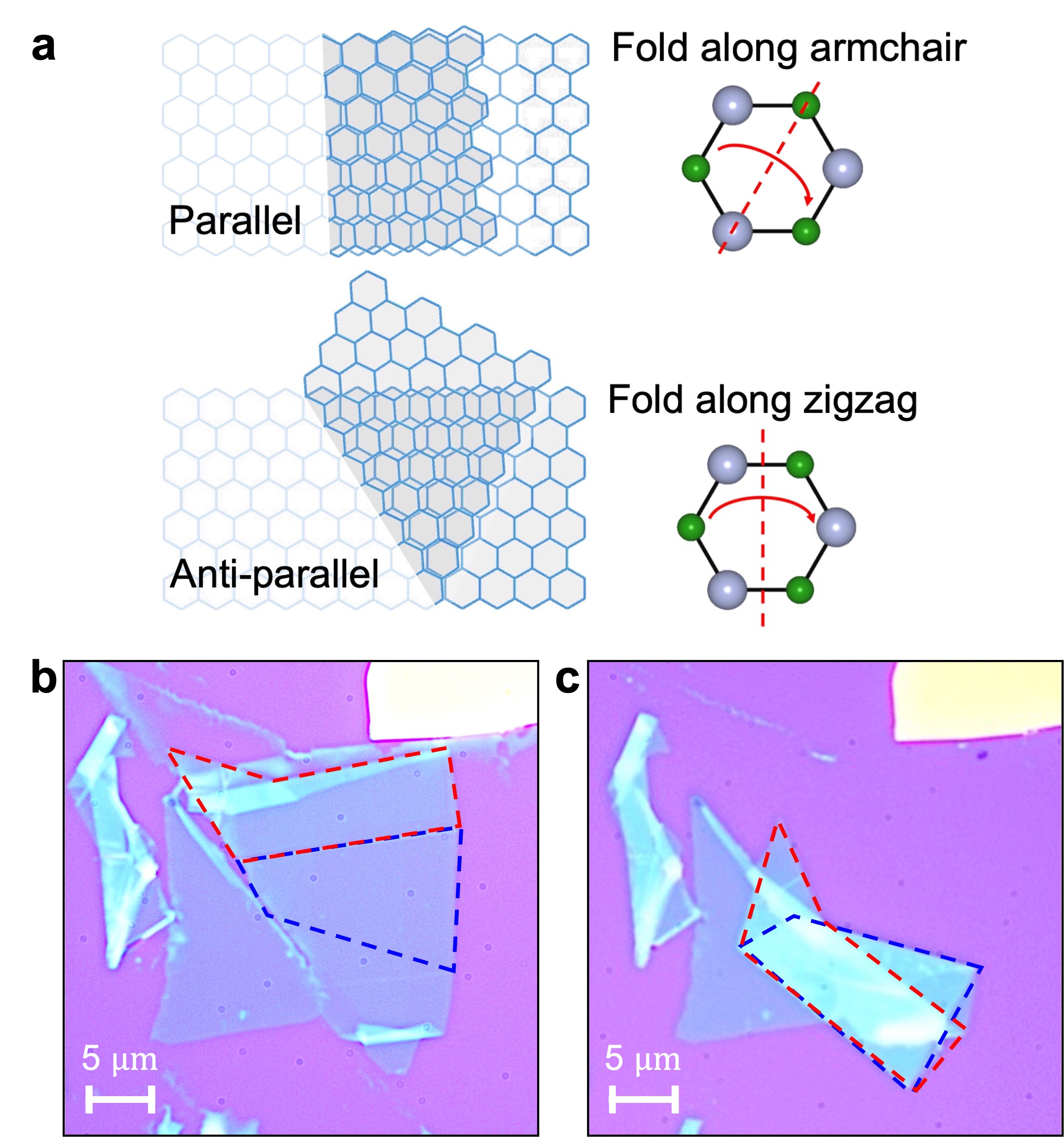}
    \caption{\textbf{Introduction of folding and folded hBN flakes.}
    \textbf{a}, Illustration of folding along the zigzag (armchair) direction leading to the parallel (anti-parallel) orientation at the interface.
    Optical image of one $\sim 4$ nm thick hBN flake (\textbf{a}) before and (\textbf{b}) after folding.
    Scale bars represent 5 $\mu$m. The folded regions are marked by dashed lines with different color contrasts. The blue and red dashed lines encircle the bilayer and trilayer regions in (\textbf{b}), respectively.}
    \label{fig:figs1}
\end{figure}

\begin{figure}[H]
    \centering
    \includegraphics[width=16cm]{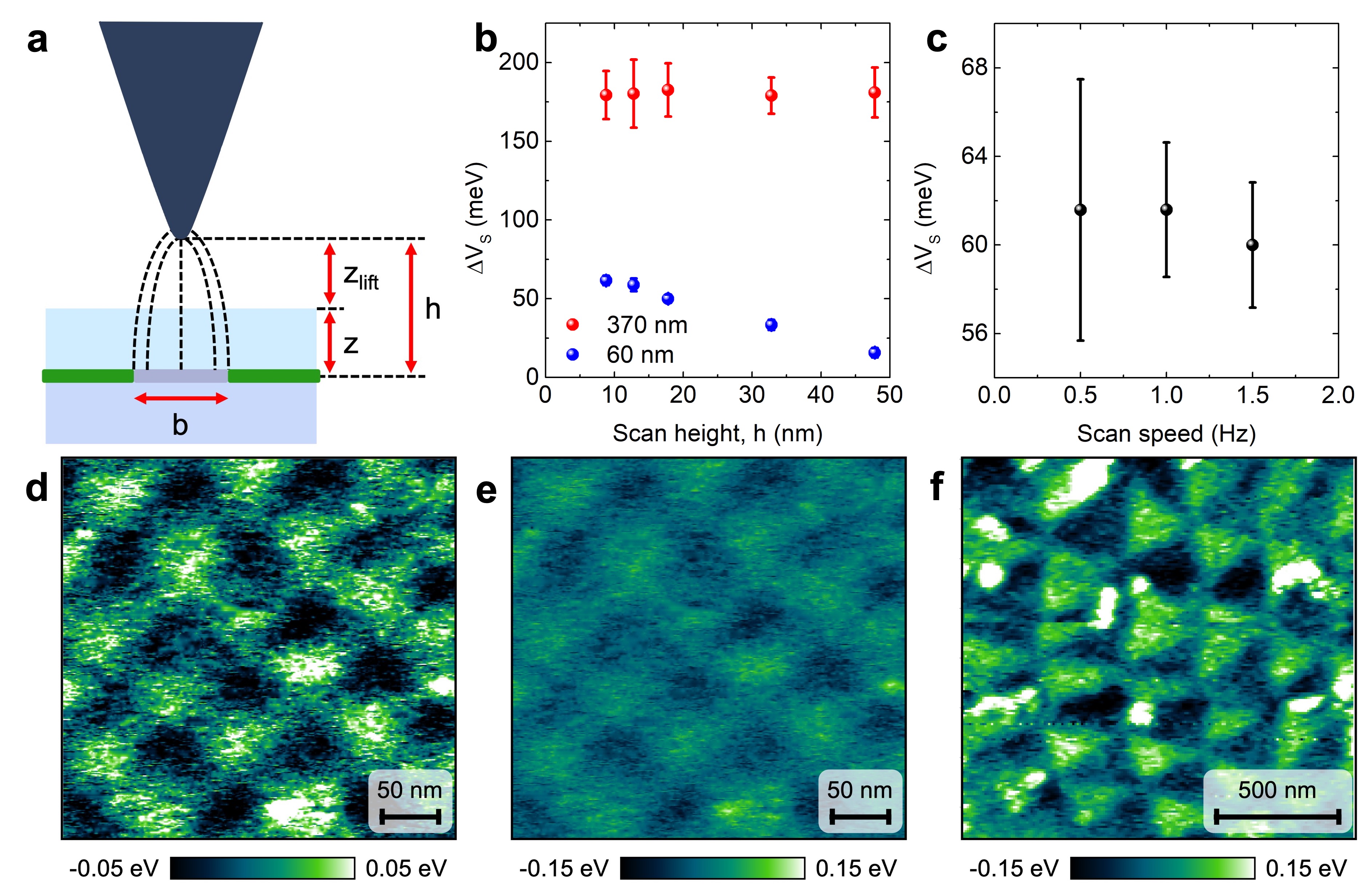}
    \caption{\textbf{Parameter dependence of KPFM measurements.}
    \textbf{a}, Illustration of the tip scanning over the sample. Scan height $h$ is the sum of the thickness of the top hBN, $z = 7.8$~nm, and tip-to-sample distance, $z_{\textrm{lift}}$. $b$ is the domain size as defined in the main text. 
    \textbf{b}, Measured modulation potential depth $\Delta V_s$ as a function of the scan height, h, for two domain sizes of 60 nm (blue sphere) and 370 nm (red sphere). The measured potential depth remains constant for large supercells while drops for small supercells with increasing scan height. It is important to note that the potential can be sensitive to scan height even for large supercells with a blunt tip and a reduced spatial resolution. The tip-to-sample distance is kept at $z_{\textrm{lift}}= 2$~nm for all results presented in the main text. 
    \textbf{c}, Potential modulation as a function of scan speed for a supercell of 60~nm period. The potential value does not change significantly for the scan speed between 0.5 and 1.5~Hz. The images presented in the main text are taken at the scan speed of 1~Hz.
    \textbf{d-f}, KPFM images of the domains with supercell size of 60~nm (\textbf{d}, \textbf{e}) and 370~nm (\textbf{f}). The images \textbf{d} and \textbf{e} are identical images but represented on two different potential scales. The potential scale of \textbf{e} is chosen to be the same as that of \textbf{f} for comparison. Notice that several small features with higher potential are clearly resolved in \textbf{d} and \textbf{e}, which is indicative of a high spatial resolution. The measurements were performed by a supersharp diamond-like carbon tip with a radius of 1 nm.}
    \label{fig:figs2}
\end{figure}

\newpage

\begin{figure}[H]
    \centering
    \includegraphics[width=11cm]{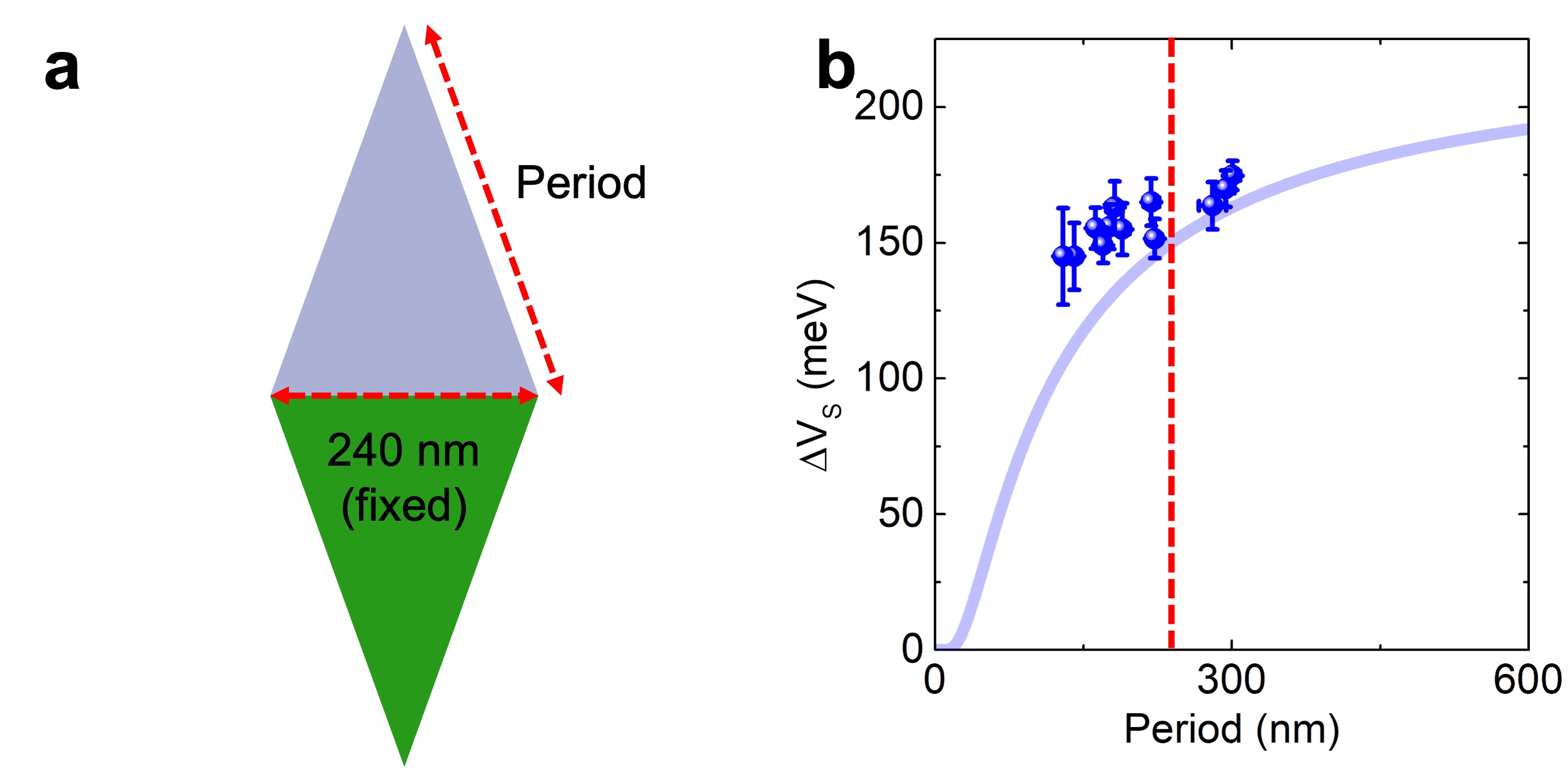}
    \caption{\textbf{Supercell shape influencing modulation potential.}
    \textbf{a}, Illustration of an isosceles triangular supercell. The base length is fixed at 240 nm and sides are defined as the period in panel b.
    \textbf{b}, The modulation potential measured from several isosceles triangular supercells. In regions where supercells are distorted by unintentional strain, equilateral triangles become isosceles triangles. In this case, the extracted potential (blue spheres) deviates from the theoretical prediction (solid curve). The vertical red dashed line corresponds to the period of 240 nm, i.e. an equilateral triangle.}
    \label{fig:figs3}
\end{figure}

\newpage

\begin{figure}[H]
    \centering
    \includegraphics[width=11cm]{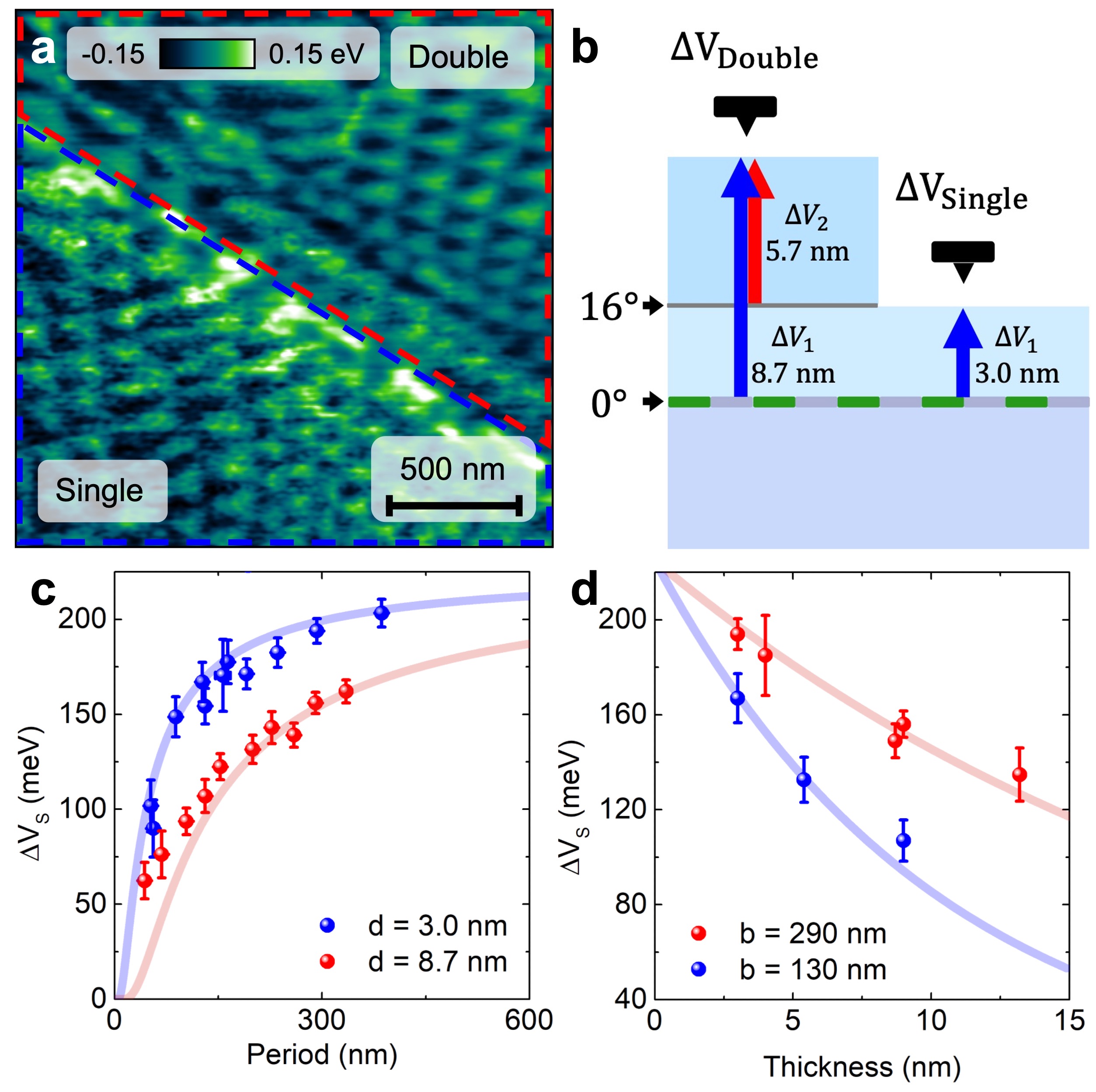}
    \caption{\textbf{Top hBN layer thickness dependent potential modulation.}
    \textbf{a}, KPFM images of single (dashed blue) and double (dashed red) moir\'e superlattices. Scale bar represents 500 nm.
    \textbf{b}, Schematic of the sample in \textbf{a}.
    \textbf{c}, Surface potentials measured from the single (blue) and double (red) moir\'e regions. The ferroelectric interface (the bottom interface) lies between layers with $\delta \sim$ 0$^{\circ}$ while the contribution from the second interface with $\delta \sim$ 16$^{\circ}$ is negligible. The shallower potential from the double moir\'e region is due to the effectively thicker hBN layer. The blue and red lines plotted the theory curve with measured hBN thickness from the bottom interface. 
    \textbf{d}, Summary of potential modulation as a function of the top hBN thickness. Data points were collected from several samples with the same supercell sizes of 130 nm and 290 nm.}
    \label{fig:figs4}
\end{figure}

\begin{figure}[H]
    \centering
    \includegraphics[width=16.5cm]{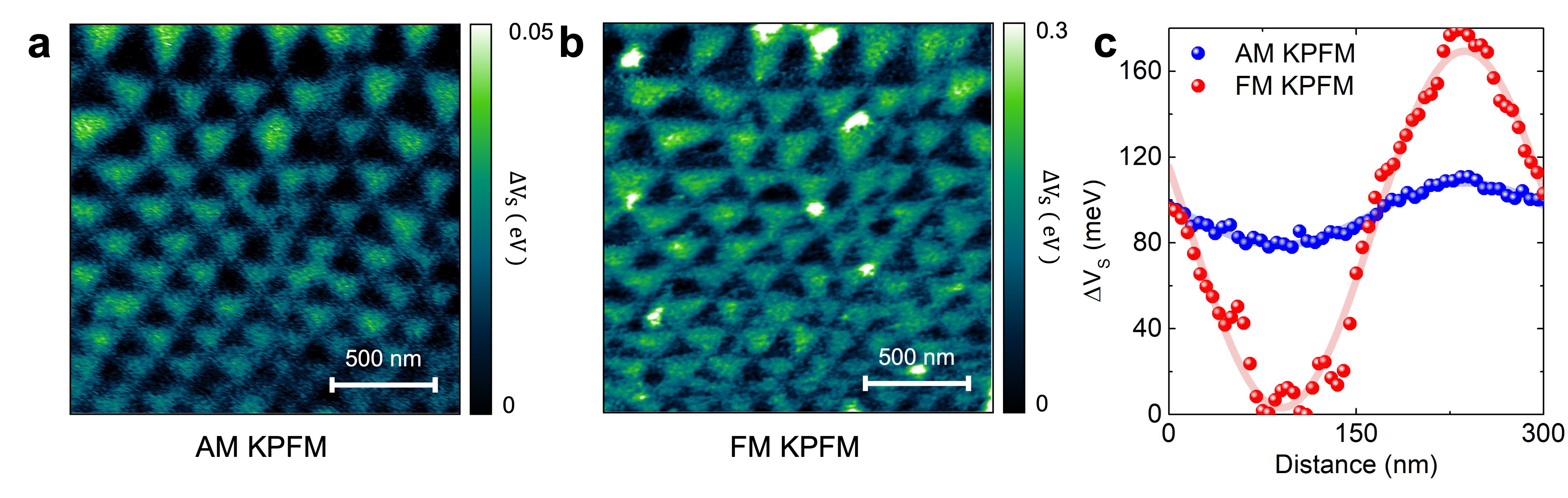}
    \caption{\textbf{Comparison between AM-KPFM and FM-KPFM.}
    \textbf{a}, AM-KPFM potential map of the sample in main text.
    \textbf{b}, FM-KPFM potential map in region.
    \textbf{c}, Surface potentials of 170 nm period moir\'e lattice measured by AM-KPFM (blue spheres) and FM-KPFM (red spheres). While AM-KPFM images show reduced noise, it is known to underestimate the potential modulation. Due to an averaging effect, AM-KPFM ($\sim$ 30 meV) yields one fifth of modulation potential extracted from FM-KPFM ($\sim$ 150 meV).}
    \label{fig:figs5}
\end{figure}

\newpage

\begin{figure}[H]
    \centering
    \includegraphics[width=16.5cm]{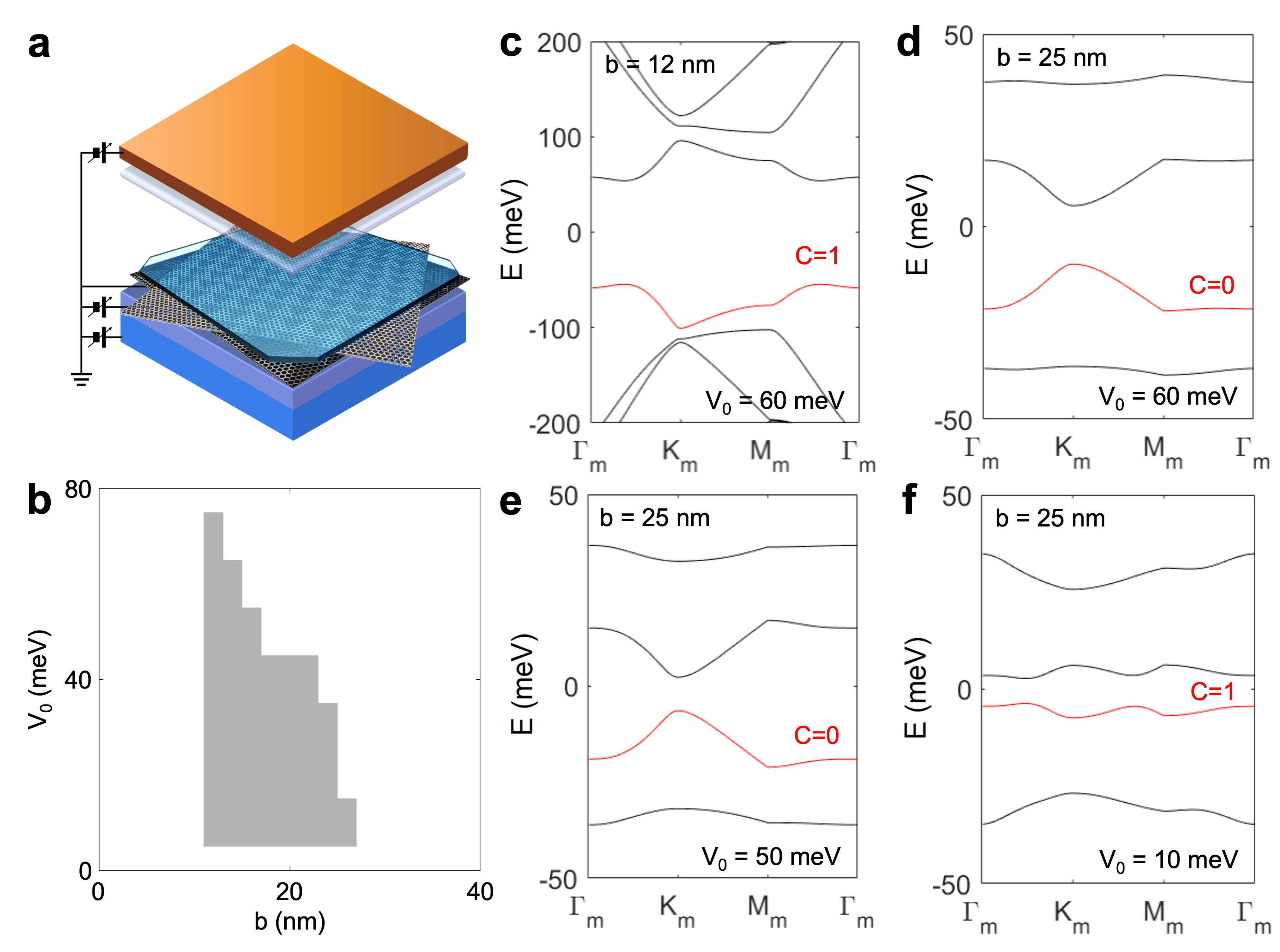}
    \caption{\textbf{Non-trivial topological states in bilayer graphene induced by twisted hBN.}
    \textbf{a}, Proposed device structure to control band topology of a natural bilayer graphene by t-hBN substrate with a gate voltage V$_0$.
    \textbf{b}, Phase diagram of the bilayer graphene under the moir\'e potential generated by twisted hBN as a function of the applied bias $V_0$ and the period of the moir\'e potential, b. Within the shaded area, the highest valence band (marked by red in \textbf{b-e}) is topologically non-trivial. Outside this area, the highest valence band is either trivial or does not have a well-defined gap with other bands.
    \textbf{c-f}, Dispersion relation of the bilayer graphene corresponding to several representative points of the phase diagram. \textbf{c}, $V_0$ = 60 meV , b = 12 nm. \textbf{d}, $V_0$ = 60 meV , b = 25 nm. \textbf{e}, $V_0$ = 50 meV , b = 25 nm. \textbf{f}, $V_0$ = 10 meV , b = 25 nm.}
    \label{fig:figs6}
\end{figure}

\newpage

\begin{figure}[H]
    \centering
    \includegraphics[width=16.5cm]{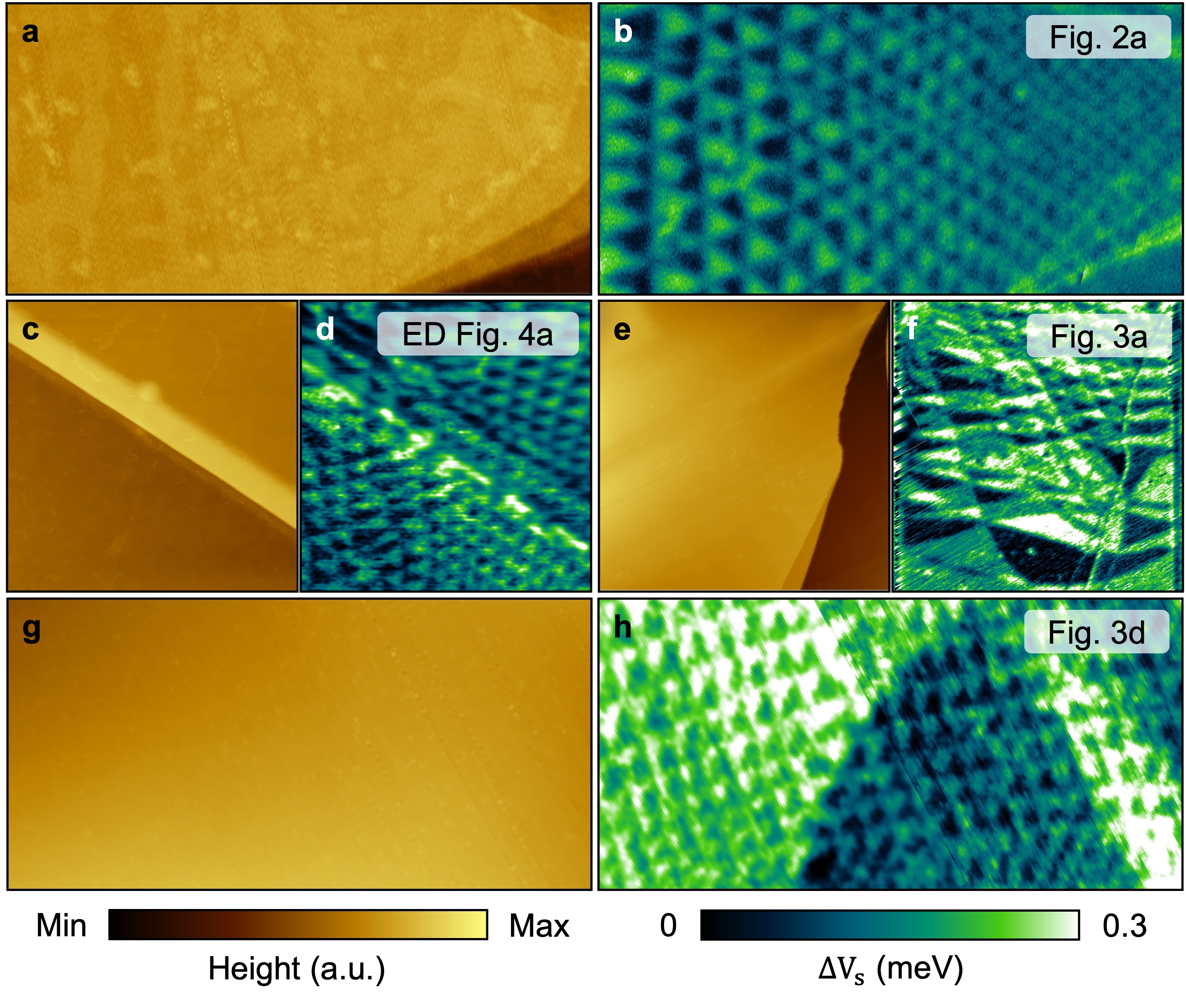}
    \caption{\textbf{Topography and KPFM images.}
    \textbf{a, c, e, g}, Topography and
    \textbf{b, d, f, h}, corresponding KPFM images for twisted bilayers and trilayers included in the main text and Extended Data. There are no correlations between these images, indicating minimal influence of topography in the modulation potential analysis.}
    \label{fig:figs7}
\end{figure}

\begin{figure}[H]
    \centering
    \includegraphics[width=16.5cm]{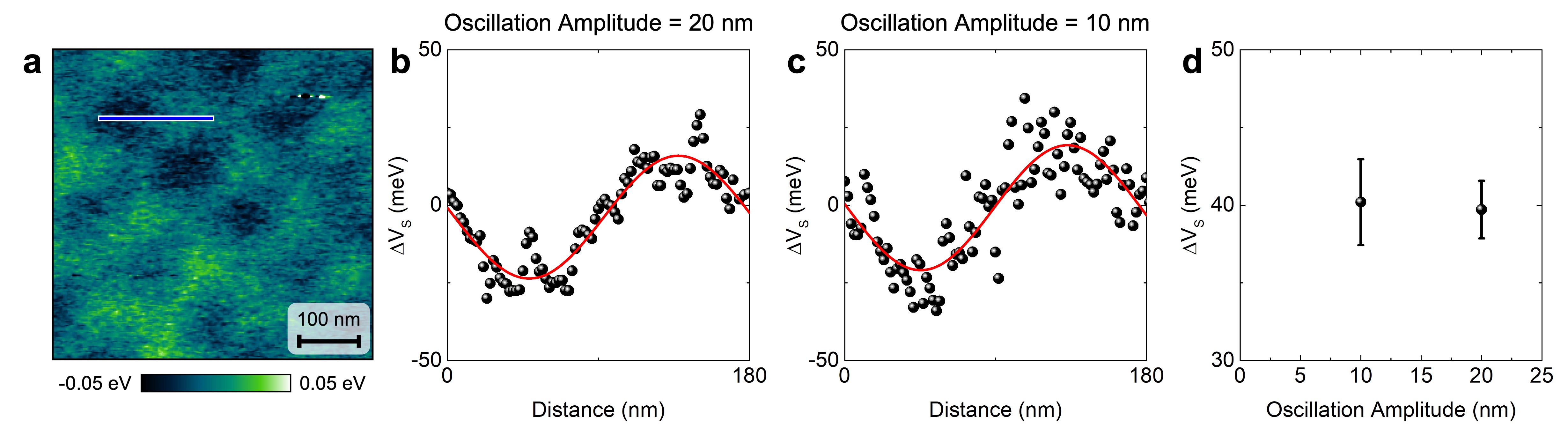}
    \caption{\textbf{Oscillation amplitude dependence of KPFM measurements.}
    \textbf{a}, KPFM potential profile with the oscillation amplitude of 20 nm for a domain size of 100 nm. The thickness of the top hBN is $z = 13$ nm, and tip-to-sample distance is, $z_{lift} = 1$ nm. The measurements were performed by a supersharp diamond-like carbon tip with a radius of 1 nm.
    \textbf{b}, KPFM potential profile with the oscillation amplitude of 10 nm for the same domain.
    \textbf{c}, Summary of modulation potential depth $\Delta$V$_S$ as a function of the oscillation amplitude. As reducing the oscillation amplitude, noise becomes significant however the potential depth is independent. With the oscillation amplitude of 5 nm, the data is too noisy so it is not meaningful to extract the potential value.}
    \label{fig:figs8}
\end{figure}

\begin{figure}[H]
    \centering
    \includegraphics[width=8cm]{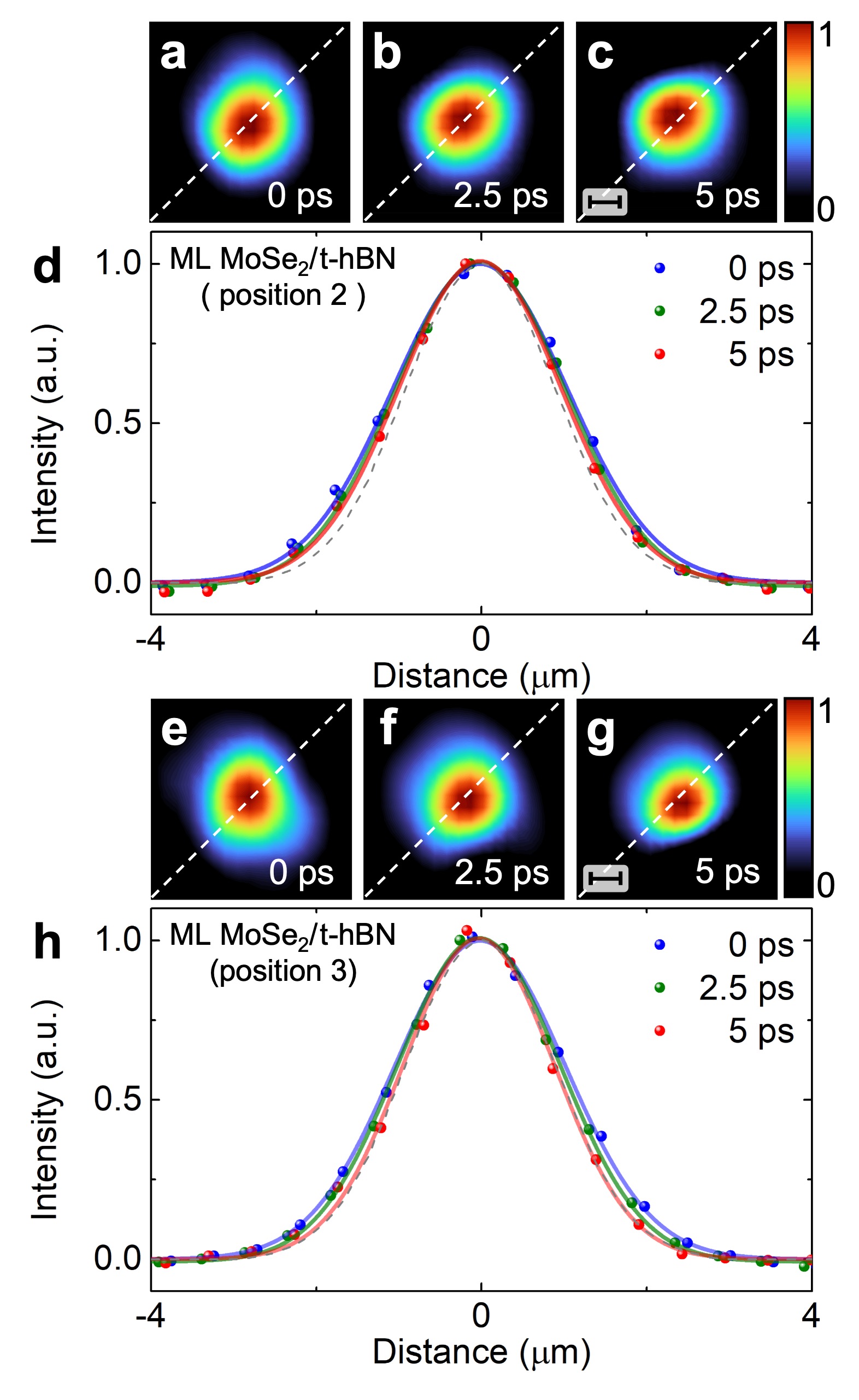}
    \caption{\textbf{Diffusion measurements in multiple locations on the ML MoSe$_2$/t-hBN.}
    \textbf{a-c}, Normalized pump-probe images from the second position on the ML MoSe$_2$/t-hBN at delay times 0 ps (\textbf{a}), 2.5 ps (\textbf{b}), and 5 ps (\textbf{c}). The scale bar is 1 $\mu$m.
    \textbf{d}, Line profiles taken along the white dashed lines in (\textbf{a-c}).
    \textbf{e-h}, Corresponding images and line profiles from the third position on the ML MoSe$_2$/t-hBN. The exciton density is the same as the main text of $\sim$ 7$\times$10$^{11}$ cm$^{-2}$.}
    \label{fig:figs9}
\end{figure}

\begin{figure}[H]
    \centering
    \includegraphics[width=8cm]{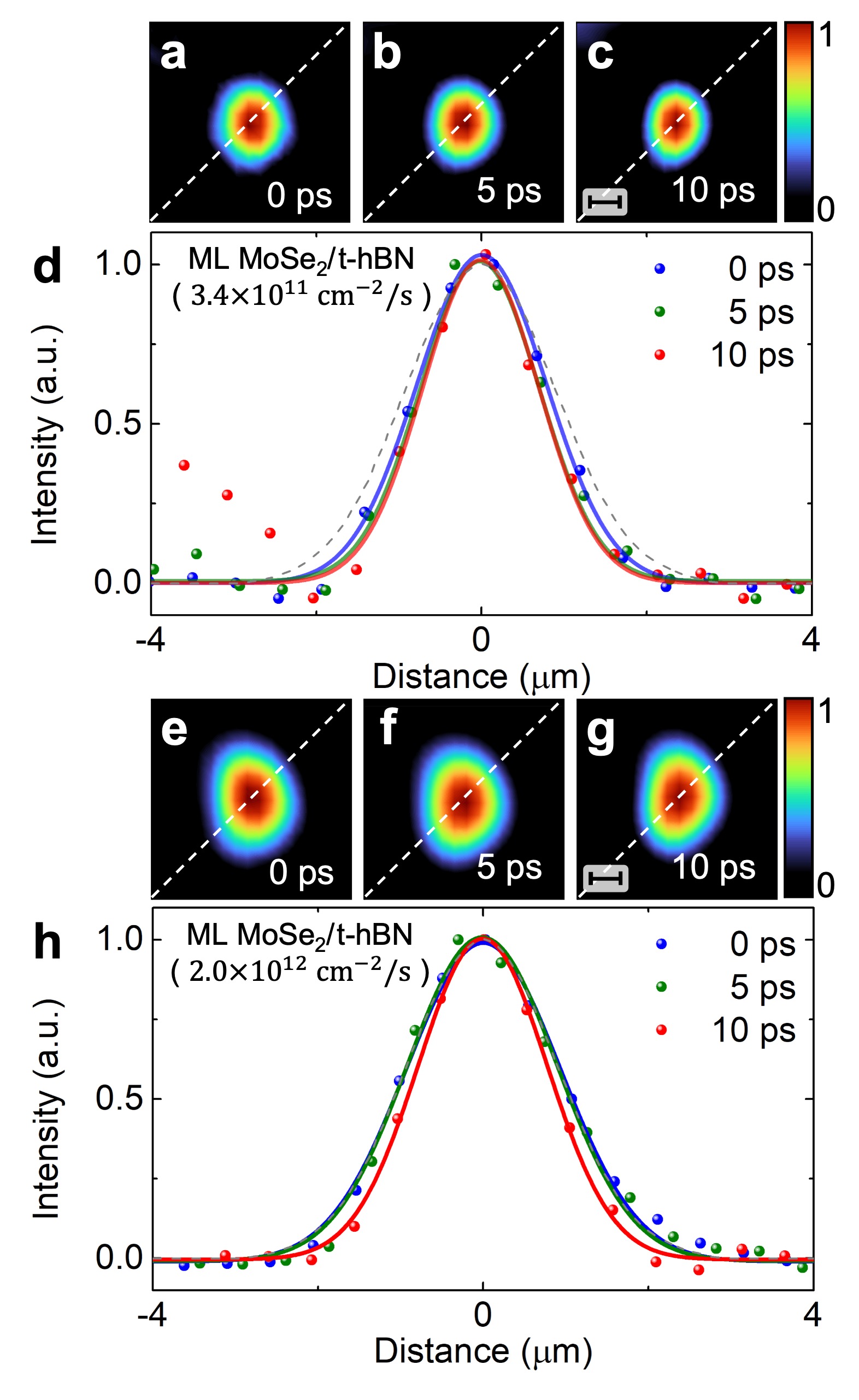}
    \caption{\textbf{Diffusion measurements with different exciton densities on the ML MoSe$_2$/t-hBN.}
    \textbf{a-c}, Normalized pump-probe images on the ML MoSe$_2$/t-hBN with exciton density of $\sim$ 3$\times$10$^{11}$ cm$^{-2}$ at delay times 0 ps (\textbf{a}), 5 ps (\textbf{b}), and 10 ps (\textbf{c}). The scale bar is 1 $\mu$m.
    \textbf{d}, Line profiles taken along the white dashed lines in (\textbf{a-c}).
    \textbf{e-h}, Corresponding images and line profiles from the ML MoSe$_2$/t-hBN substrate with exciton density of $\sim$ 2$\times$10$^{12}$ cm$^{-2}$.}
    \label{fig:figs10}
\end{figure}

\bibdata{hBN-KPFM.bib}

\end{document}